\documentstyle[12pt,epsfig]{article}
\newfont{\cmu}{cmu10 scaled\magstep1}
\newcommand{\be}{\begin{equation}}
\newcommand{\ee}{\end{equation}}
\newcommand{\bea}{\begin{eqnarray}}
\newcommand{\eea}{\end{eqnarray}}

\begin{document}
\vspace*{2cm}
\begin{center}
{\Large\bf Impact parameter dependencies in Pb(160 AGeV)+Pb reactions --
hydrodynamical vs. cascade calculations}
\\[.5cm]
{\bf J. Brachmann$^1$, M. Reiter$^1$,  M. Bleicher$^1$, A. Dumitru$^2$,
J.A. Maruhn$^1$, H. St\"ocker$^1$, W. Greiner$^1$}
\\[0.2cm]
{\small ${}^1$ Institut f\"ur Theoretische Physik, Universit\"at
Frankfurt a.M., Germany}
\\[0.4cm]
{\small ${}^2$ Department of Physics, Yale University, New Haven,
Connecticut, USA}
\\[1cm]
{January 18, 1999}
\end{center}

\begin{figure}[h]
\centerline{\hbox{\psfig{figure=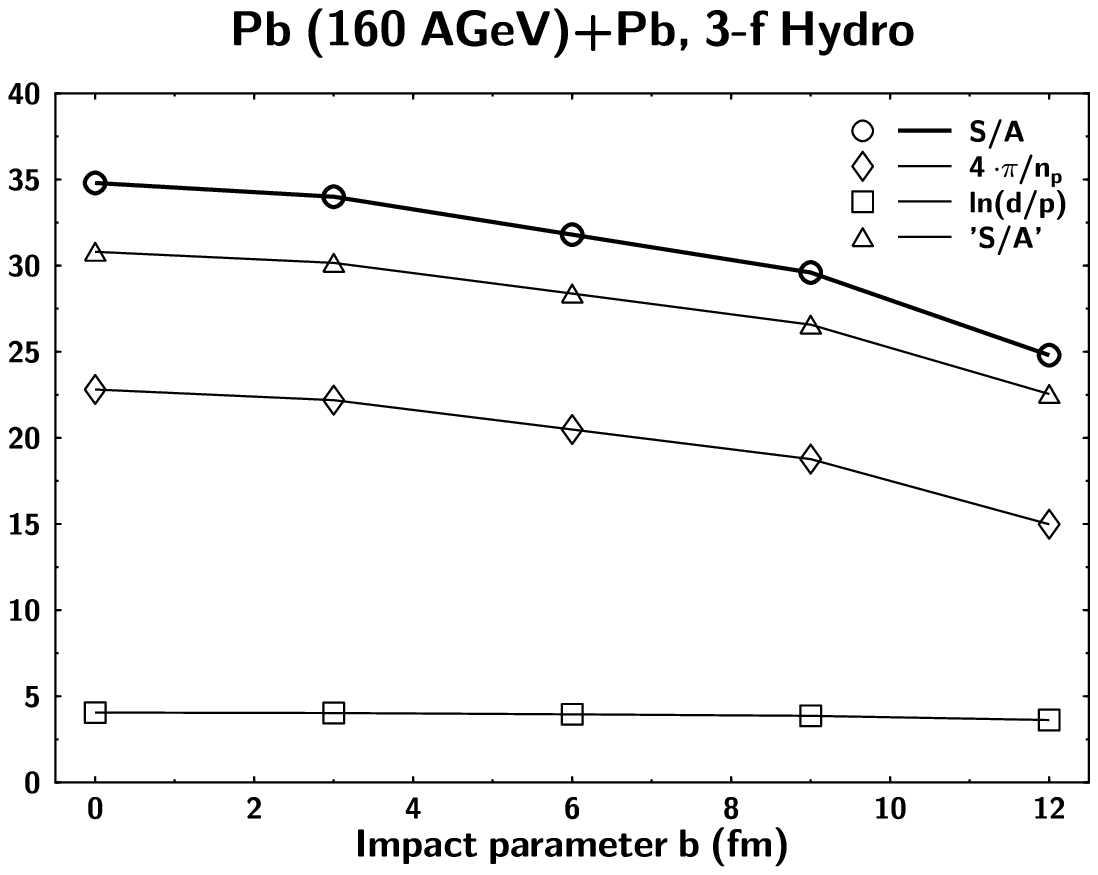,width=10cm}}}
\caption{Specific Entropy $S/A$ vs. Impact parameter $b$. Circles
denote the $S/A$ values from the 3-fluid model. Diamonds and
squares denote the ratios of pions to baryons and deuterons to protons
(logarithmic), resp., as calculated from  the $S/A$ values.
Triangles show the
\lq$S/A$\rq ~values parametrized by
$S/A=3.945+ln(d/p)+4(\pi/n_p)$.}   
\label{sa}
\end{figure}

Particle ratios are an appropriate tool to study the characteristics
of entropy production in heavy-ion collisions, as shown in
Fig.~\ref{sa}.
We investigate the impact parameter dependence of the $S/A$ ratio
(entropy
$S$ per net participating baryon $A$) by means of the
$\overline\Lambda /\overline p$  ratio and will find it a tool to
distinguish
between chemical equilibrium as assumed in hydrodynamics (here: the
3-fluid model \cite{3f}) and chemical
non-equilibrium like in microscopic models as the UrQMD model
\cite{urqmd}.

\begin{figure}[h]
\centerline{\hbox{\psfig{figure=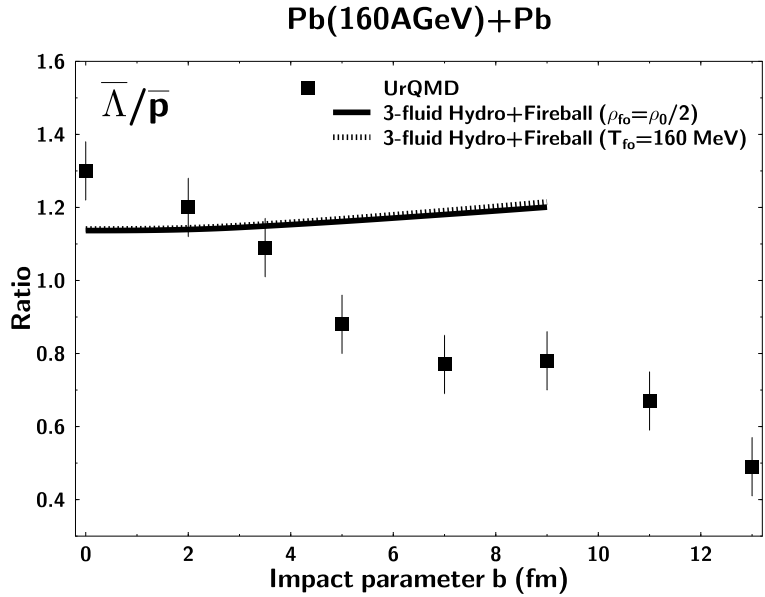,width=10cm}}}
\caption{Impact parameter dependence of the $\overline\Lambda
/\overline p$ ratio in Pb(160~AGeV)+Pb reactions \protect\cite{alap}. The full
squares
denote the UrQMD calculations, the lines show the 3-Fluid Hydro
+ Fireball calculation including a first order phase transition (full
line: freeze-out at $\rho=\rho_0/2$, dashed line: freeze-out at
$T=160MeV$).}
\label{Falap}
\end{figure}

In the 3-fluid hydrodynamical model \cite{3f,Manuel-S/A}
an EoS with a first order phase transition to a QGP is used.
We employ that model to calculate $S/A$ during the initial  
stage of the reaction as a function of impact parameter b   
(cf.\ \cite{Manuel-S/A}), as shown in Fig. \ref{Falap}.      
To show the behaviour of the $\overline\Lambda /\overline p$  ratio
for the chemical equilibrium case, the creation of a fireball, composed
of all
hadrons up to mass $m=2$~GeV, with a uniform $S/A$ ratio
and net baryon density $\rho$ is assumed.
The hadron ratios are calculated assuming
chemical freeze-out at a net baryon density $\rho=\rho_0/2$,
$T=160$~MeV,
resp.~.

Within the UrQMD model \cite{urqmd}, the
non-equilibrium dynamics are treated in a microscopic hadronic scenario.
Baryon-baryon, meson-baryon and meson-meson collisions
lead to the formation and decay of resonances and color flux tubes. The
produced, as well as the incoming particles,
rescatter in the further evolution of the system.

Fig. \ref{Falap} shows the $\overline\Lambda /\overline p$ ratio for
different impact pa\-ra\-me\-ters \cite{alap}.
In the 3-fluid  approach the $\overline\Lambda /\overline p$
ratio stays constant with b.
In contrast, the hadronic UrQMD model yields a strong
dependence of this ratio on impact parameter b.
The $\overline\Lambda/\overline p$ ratio drops rapidly with
increasing b from 1.3 to 0.5.

In the microscopic UrQMD model, there is an interplay
between particle production and subsequent annihilation.
In peripheral collisions the $\overline\Lambda$ production is
basically the same as in p+p reactions.
Due to the mass difference  between strange and up/down quarks the
production of (anti-)strange quarks is suppressed, which
results in a suppression of $\overline\Lambda$ over
$\overline p$ by a factor of 2.
In central Pb+Pb encounters meson-baryon and meson-meson interactions
work
as additional sources for the anti-hyperon and anti-proton production
and additional rescattering has
to be taken into account in the hot and dense medium. Anti-baryons are
strongly affected by the comoving baryon density and annihilate, while,
according to the additive quark model,
the annihilation probability for $\overline\Lambda$'s
is smaller,
leading to an increase of the
$\overline\Lambda /\overline p$ ratio above 1.
Thus, in the UrQMD \cite{urqmd} chemical equilibrium may only be
established in
very central reactions, where enough secondary collisions drive the
system into
chemical equilibrium.
In contrast, the fluids of the hydrodynamical calculation are, by
definition,
in chemical\footnote{In the beginning of the reaction kinetic
equilibrium between the fluids is not assumed, but chemical equilibrium
is established by assumption of an EoS.} equilibrium.

\end{document}